\documentclass[aps,twocolumn]{revtex4-1}

\usepackage{amsmath}

\pdfoutput=1
\usepackage{graphicx}

\begin{document}

\title{Modeling cochlear two-tone suppression using a system of nonlinear oscillators\\with feed-forward coupling}

\author{Masanori Ouchi and Hiroya Nakao}

\address{Department of Systems and Control Engineering, Tokyo Institute of Technology, O-okayama 2-12-1, Meguro, Tokyo 152-8552, Japan}

\begin{abstract} 
  Mechanism of two-tone suppression is studied using
  a coupled-oscillator model of the cochlea with feed-forward coupling.
  Local amplification of sound signals is modeled by using Stuart-Landau
  oscillators near the Hopf bifurcation, and transmission of sound signals
  is described as feed-forward coupling between the oscillators.
  Effect of suppressor signals on the response to probe signals is analyzed by numerical simulations.
  It is found that the effect of suppression is qualitatively different
  depending on relative frequency between probe and suppressor signals.
  By analyzing a simplified two-oscillator model, we explain the mechanism
  of the suppression, where configuration of the oscillators plays an essential role.
\end{abstract}

\maketitle

\section{Introduction}  

The cochlea in the inner ear is an auditory sensory organ that transforms sound stimuli into neural signals. It is known that the cochlea has an active amplification mechanism, which realizes sharp frequency selectivity and a wide dynamic range~\cite{manley2000,hudspeth2008,kohda1985,eguilz2000,camalet2000,magnasco2003,hudspeth2014}. 
The cochlear duct has a tubular shape, which is separated by a basilar membrane and filled with lymphatic fluid. Sound stimuli coming from the eardrum propagate through this fluid as pressure waves, and variations in the fluid pressure induce vibrations of the basilar membrane. Hair cells attached to the basilar membrane 
actively amplify the mechanical vibrations and transform them into neural signals.

Depending on the frequency of the sound stimulus, 
active vibrations of the hair cells are evoked at different points on the basilar membrane.
The most sensitive frequency at each point of the basilar membrane is called the characteristic frequency (CF). The CF is exponentially distributed along the cochlea, from higher frequencies at the base (near the entrance) to lower frequencies at the apex (near the end of the duct). In human, the CF ranges approximately from 20Hz (apex) to 20kHz (base).

It has been shown that the vibrations of the basilar membrane and the hair cells can be modeled by using a Stuart-Landau oscillator, a normal form of the Hopf bifurcation~\cite{kohda1985,eguilz2000,camalet2000,magnasco2003,hudspeth2014}. It is considered that the
cochlea can be modeled as an array of such active oscillators slightly below the onset of spontaneous oscillation, which are coupled mechanically and through the lymphatic fluid with high viscosity in the cochlear duct~\cite{kern2003,duke2003,dierkes2008,meaud2011,wit2012,Fruth2014,gomez2016}.
Development of biomimetic acoustic sensors that take into account the amplification characteristics of the cochlea has also been attempted~\cite{kohda1985,martignoli2007,gomez2014,Torikai2015}.

In this paper, we study the effect of coupling on two-tone suppression, a well-known auditory phenomenon in which response to a probe signal is reduced when another suppressor signal with different frequency is presented~\cite{Rhode1993}.
Depending on whether the frequency of the suppressor signal is higher or lower than that of the probe signal, it is classified into high-side suppression (HSS) and low-side suppression (LSS), which show qualitatively different characteristics.
By numerical simulations and theoretical analysis of the coupled-oscillator model, we analyze how the effect of suppression depends on the level and frequency of the suppressor signal, and argue that the qualitative difference between HSS and LSS is caused by the difference in relative configuration of the oscillators.

\vspace*{-3mm}
\section{Model}

In this study, we model the propagation of sound pressure in the cochlea using coupled oscillators with feed-forward coupling.
Each oscillator represents vibrations of the basilar membrane and the attached hair cells, and the feed-forward coupling is assumed
to represent the unidirectional propagation of sound waves in the lymphatic fluid from the base to the apex observed experimentally~\cite{pickles2012}.
Similar models have also been considered in Refs.~\cite{martignoli2007,gomez2014}.

\begin{figure}[b]
\centering
    \includegraphics[width=1\linewidth,clip]{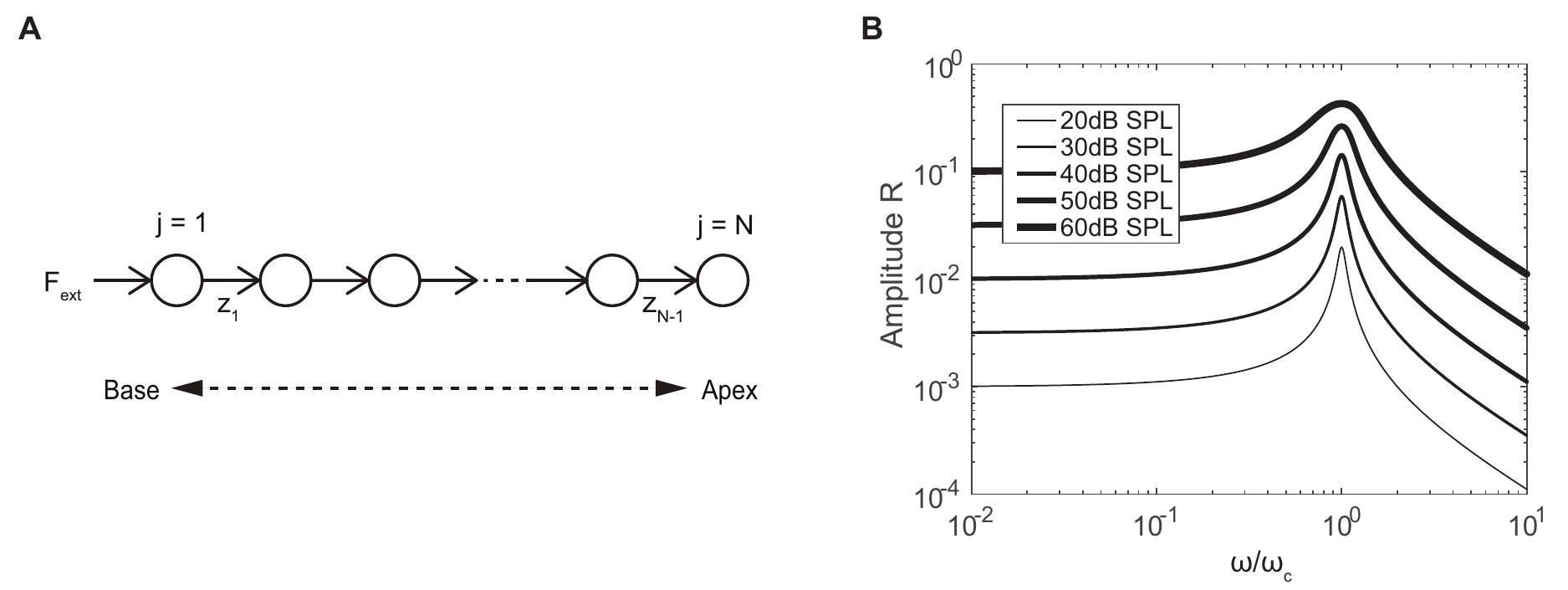}
  \caption{(A) Schematic illustration of the model. The basilar membrane is modeled as a one-dimensional array of $N$ oscillators with feed-forward coupling. The external input unidirectionally propagates from the base ($j=1$) to the apex ($j=N$). (B) Amplitude response of a single oscillator to a sinusoidal external input vs. relative frequency of the input signal to the natural frequency of the oscillator. Results for several values of the input amplitude are shown.}
  \label{fig1}
\end{figure}

Figure~\ref{fig1}A shows a schematic illustration of the model, a one-dimensional array of nonlinear oscillators with feed-forward coupling.
Each oscillator is described by a Stuart-Landau model, whose characteristic frequency gradually varies along the cochlea, and receives an input from the previous oscillator.
The dynamics of the model is described by
\begin{align}
  \dot{z_j} = \omega_{c,j} [ (\mu_j+i) z_j - |z_j|^2 z_j + F_j ]
  \label{model}
\end{align}
for $j=1, 2, ..., N$, where $i=\sqrt{-1}$, $z_j(t)$ is a complex variable representing the state of the oscillator $j$ at time $t$, $\omega_{c, j}$ is its natural frequency, $\mu_j$ is a bifurcation parameter, and $F_j(t)$ is an input from the previous oscillator.

The real part of $z_j(t)$ corresponds to the displacement of the basilar membrane. 
The bifurcation parameter $\mu$ takes a negative value close to zero, which represents that the oscillators are slightly below the critical point of the Hopf bifurcation. Physiologically, this parameter characterizes dynamical properties of the hair cells and basilar membrane, as well as the viscous lymphatic fluid, and represents how close the system is to the onset of spontaneous oscillation.
We assume that the input is given by $F_j(t) = z_{j-1}(t)$ for $j=2, ..., N$, namely, the oscillation of the previous oscillator is directly relayed to the next oscillator without transmission loss.

The oscillator $j=1$ at the base of the cochlea receives an external input $F_{ext}$ from the eardrum.
We consider two-tone stimuli and assume that the external input is a superposition of a probe signal with frequency $\omega_{probe}$ and a suppressor signal with frequency $\omega_{sup}$, both of which are sinusoidal.
The external input is thus given by
\begin{align}
  F_{ext}(t) = f e^{i\omega_{probe} t} + g e^{i\omega_{sup} t},
\end{align}
where $f \geq 0$ and $g \geq 0$ are the amplitudes of the probe and suppressor signals, respectively.

We use $N=10$ oscillators and fix their bifurcation parameters at $\mu_j = \mu = -0.05$ for all $j=1, ..., N$. This value is slightly below the critical value $\mu_c = 0$ of a Hopf bifurcation.
The properties of the system are qualitatively the same for other values of $\mu$, as long as $\mu$ takes a negative value close to zero.
Natural frequencies of the oscillators are assumed to be $ \omega_{c,j}=\omega_{c,1}\,\gamma^{-(j-1)} $, taking into account the exponential distribution the CF in the actual cochlea~\cite{pickles2012}.
Ratio of natural frequencies between two neighboring oscillators is set at $ \gamma=\omega_{c,n}/\omega_{c,n-1}=2 \ (n=2, \cdots, N)$, and the natural frequency of the first oscillator is fixed at $ \omega_{c,1}=10^5 \mbox{[rad/s]}$.

With these parameters, each oscillator converges to a stable fixed point at $z_j = 0$ when no input is given ($F_j = 0$). When the oscillator receives a periodic input whose frequency is close to its natural frequency $\omega_{c, j}$, it starts to exhibit a stable limit-cycle oscillation with the same frequency as the periodic input, thereby actively amplifying the input signal. When the input frequency is lower than the natural frequency, the oscillator does not exhibit a significant response and conveys the input signal to the next oscillator without amplification or attenuation. On the other hand, when the input frequency is higher than the natural frequency, the oscillator does not respond actively and the signal is conveyed to the next oscillator after some attenuation.
In the next section, we perform numerical simulations of the model Eq.~(\ref{model}) to analyze its response properties to two-tone signals.

Before going into numerical simulations, it is instructive to see the response properties of a single oscillator, following Egu{\'\i}luz {\it et al.}~\cite{eguilz2000}. Figure~\ref{fig1}B shows the response of a single Stuart-Landau oscillator to a monotone sinusoidal input, $F(t) = a e^{i \omega t}$, where the response amplitude $R = |z(t)|$ (constant for a sinusoidal input) is plotted as a function of the relative frequency $\omega / \omega_c$ for several values of the amplitude $a$ of the input signal. These curves depend only on the relative frequency $\omega / \omega_c$ and are independent of the absolute value of the natural frequency $\omega_c$.
Here, the intensity of a sound signal is characterized by the sound pressure level (SPL), where $0\mathrm{dB\,SPL}$ corresponds to the amplitude of $10^{-4}$.
20dB increase in the SPL corresponds to 10 times increase in the amplitude of the sound signal.
The oscillator exhibits the maximal response to the input signal with $\omega = \omega_c$, which is sharply amplified when $a$ is small and compressed when $a$ is large.
It is known that these frequency selectivity and nonlinear amplification and compression are close to those of the actual basilar membrane~\cite{eguilz2000}. 
In the following, we identify the CF of the basilar membrane with the natural frequency $\omega_{c}$ of the corresponding oscillator.

\begin{figure}[t]
\centering
    \includegraphics[width=1\linewidth,clip]{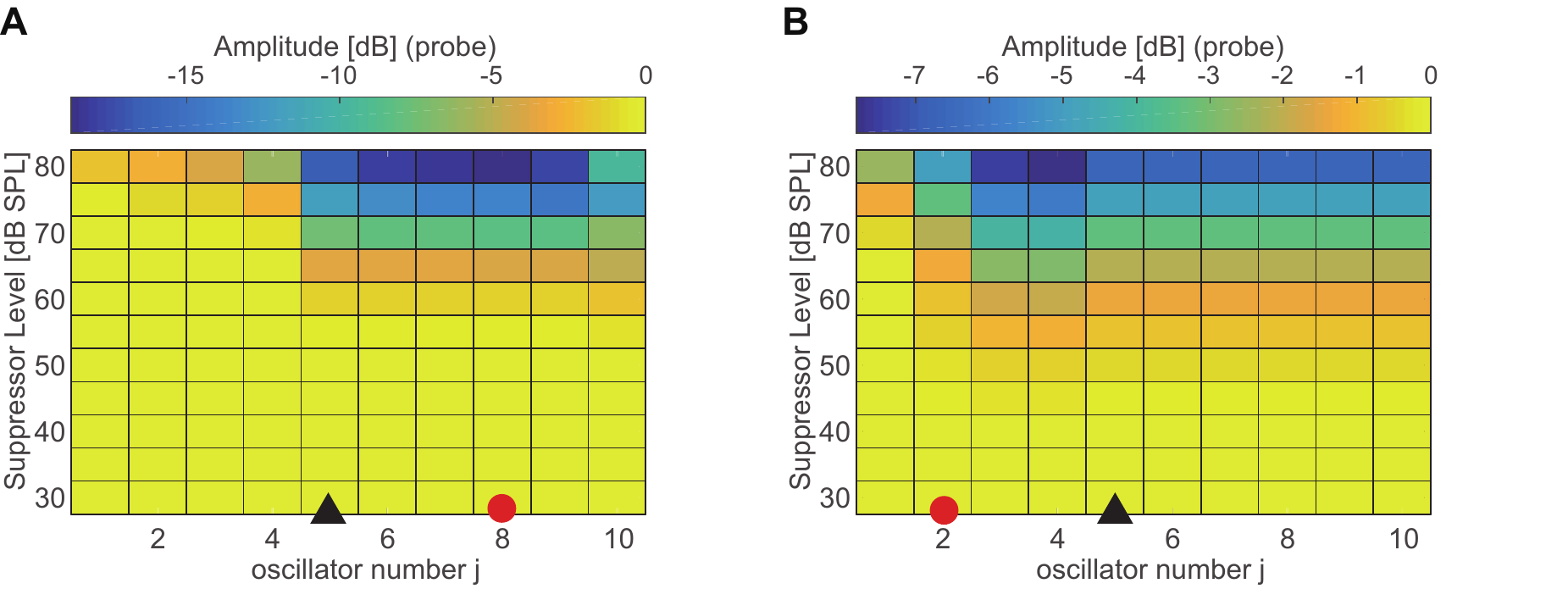}
  \caption{Suppression of response amplitude to probe signals due to suppressor signals. In each figure, the black triangle indicates the oscillator with $\mbox{CF}_{probe}$ and the red circle indicates the oscillator with $\mbox{CF}_{sup}$. Logarithm of the relative response amplitude of each oscillator is plotted in color scale. (A) Low-side suppression (LSS) with $ \Delta \omega_{sup}=0.125 $. (B) High-side suppression (HSS) with $ \Delta \omega_{sup}=8 $.  }. 
  \label{fig2}
\end{figure}

\vspace*{-3mm}
\section{Numerical simulations} 

To analyze the response properties of the model to two-tone stimuli, we perform direct numerical simulations of Eq.~(\ref{model}). We fix the frequency and level of the probe signal at $ \omega_{probe}/2\pi=994.7\mathrm{Hz}$ and $30\mathrm{dB\,SPL}$, respectively, and vary the frequency and level of the suppressor signal. The suppressor frequency is specified by its ratio to the probe frequency, $  \Delta \omega_{sup}=\omega_{sup}/\omega_{probe}$.
The output of each oscillator can approximately be represented as a superposition of the two main frequencies as $z_j(t) \approx A_j e^{i \omega_{probe} t} + B_j e^{i \omega_{sup} t}$ when $\omega_{probe}$ and $\omega_{sup}$ are not too close, where $A_j$ and $B_j$ are response amplitudes of the oscillator $j$ to probe and suppressor signals, respectively.

Figure~\ref{fig2} shows the change in the response amplitude of the oscillator to the probe signal caused by the suppressor signal, where logarithm of the relative response amplitude of each oscillator to the probe signal, $ 10\log_{10}(A_j/A_{j, 30{\rm dB}}) $, is plotted in color scale with respect to the oscillator number and the suppressor level. Here, $A_{j, 30{\rm dB}}$ is a reference response amplitude to the probe signal when a suppressor signal of 30{\rm dB} SPL is applied. 
Figure~\ref{fig2}A is for the LSS case with $ \Delta \omega_{sup}=\omega_{sup}/\omega_{probe}=0.125 $, i.e., when the suppressor frequency is lower. Similarly, Fig.~\ref{fig2}B is for the HSS case with $\Delta \omega_{probe} =8 $, i.e., when the suppressor frequency is higher. In both figures, the oscillator $j=5$ has a CF that is equal to the probe signal (hereafter denoted as $\mbox{CF}_{probe}$).  

In the LSS case, we can observe that the effect of suppression is stronger for the oscillators with $j \geq 6$ behind the probe oscillator $j=5$ with $\mbox{CF}_{probe}$, that is, for the oscillators having lower CFs than the probe frequency.
In contrast, in the HSS case, the effect of suppression is stronger for the oscillators with $j \leq 4$ in front of the probe oscillator $j=5$ with $\mbox{CF}_{probe}$, namely, for the oscillators whose CFs are higher than the probe frequency.

The effect of the suppressor signal on the oscillator $j=5$ with $\mbox{CF}_{probe}$ is physiologically important, because the sound pressure is detected around such a point whose CF is close to the probe frequency in the actual basilar membrane.
We thus analyze dependence of the effect of suppression on the frequency and level of the suppressor signal at this oscillator.

\begin{figure}[tb]
\centering
    \includegraphics[width=1\linewidth,clip]{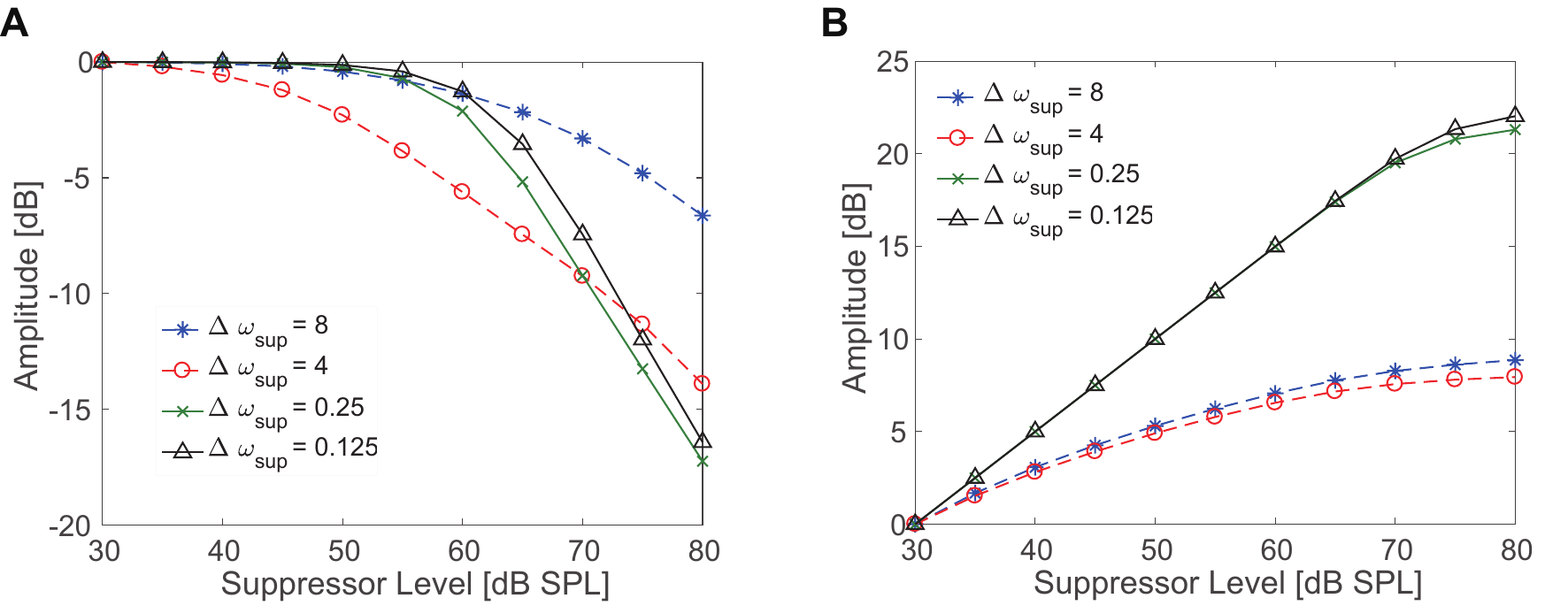}
   \caption{(A) Response amplitude of oscillator $j=5$ to the probe signal vs. suppressor level. (B) Response amplitude of the oscillator $j=5$ to the suppressor signal vs. suppressor level. In each graph, logarithm of relative amplitude to the probe or suppressor signal is plotted. The curves with $ \Delta \omega_{sup}=8, 4$ are for the HSS case, and those with $ \Delta \omega_{sup}=0.25, 0.125$ are for the LSS case.}
  \label{fig3}
\end{figure}

Figures~\ref{fig3}A and B show the response amplitudes of the oscillator $j=5$ to the probe signal and to the suppressor signal, respectively, for the cases with $\Delta \omega_{sup}=8,4$ (HSS) and $ \Delta \omega_{sup}=0.25, 0.125$ (LSS) as functions of the suppressor level.
As in Fig.~\ref{fig2}, logarithm of relative response amplitudes to the probe and suppressor signals, $10 \log_{10} (A_j / A_{j, 30{\rm dB}})$ and $10 \log_{10} (B_j / B_{j, 30{\rm dB}})$, are plotted,
where $A_{j, 30{\rm dB}}$ and $B_{j, 30{\rm dB}}$ are the reference response amplitudes to the probe and suppressor signals when a suppressor signal of 30{\rm dB} SPL is applied.

It can be seen that the effect of suppression is stronger when the probe frequency and suppressor frequency are closer for both LSS and HSS.
Here, it is notable in Fig.~\ref{fig3}A that the decay of the curves for HSS is considerably slower than those for LSS. Thus, there is a qualitative difference in the effect of suppression between LSS and HSS. Such an asymmetry is also observed experimentally in the actual cochlea~\cite{Rhode1993}. 
The response amplitude to the suppressor signal also exhibits qualitatively different dependence on the suppressor level between LSS and HSS as shown in Fig.~\ref{fig3}B. 

\begin{figure}[b]
\centering
    \includegraphics[width=0.9\linewidth,clip]{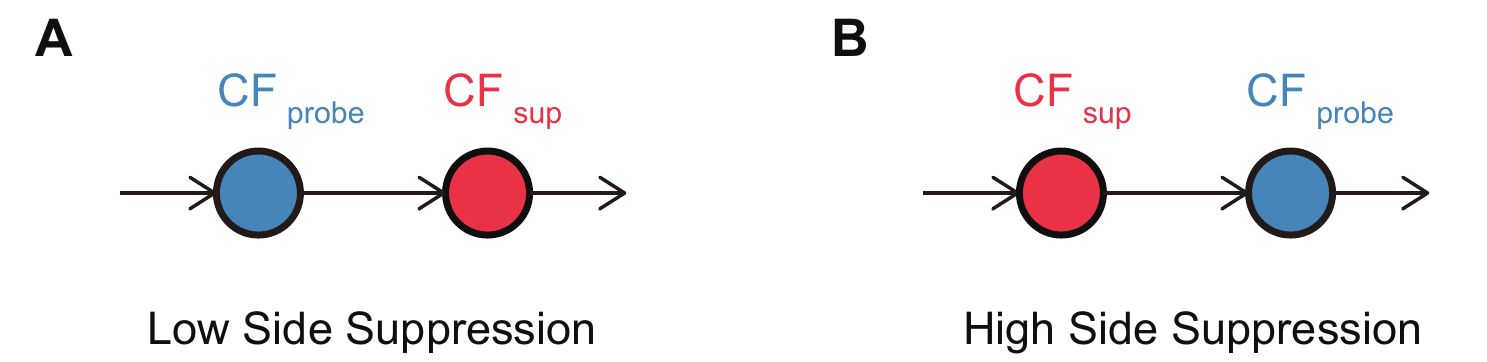}
  \caption{Simplified model with two oscillators. The natural frequency of each oscillator is equal to either of the probe or suppressor frequency. (A) Low-side suppression (LSS). (B) High-side suppression (HSS).}
  \label{fig4}
\end{figure}

\vspace*{-3mm}
\section{Theoretical analysis}

In the previous section, we have observed that the effect of suppression is qualitatively different between LSS and HSS by numerical simulations. In this section, we theoretically analyze a simplified model to clarify the dependence of the effect of suppression on the suppressor frequency and level, and explain the origin of the asymmetry between LSS and HSS observed in the numerical simulations.

From the one-dimensional structure of the model and the response property of a single oscillator in Fig.~\ref{fig1}B, propagation of the input signal in the model can be considered as follows. The external input received at the base propagates along the oscillators without significant amplification or attenuation until it reaches the oscillator whose natural frequency $\omega_{c}$ is close to the input frequency $\omega$. The signal is then selectively and nonlinearly amplified or compressed at this oscillator, and further propagated toward the apex with gradual attenuation.

Thus, it is expected that the essential difference between LSS and HSS is whether the oscillator with $\mbox{CF}_{sup}$, whose CF is equal to the suppression frequency $\omega_{sup}$, is located in front of the probe oscillator with $\mbox{CF}_{probe}$ or behind it.
To understand the consequence of this difference, we analyze a simplified model with just two oscillators whose characteristic frequencies are $\mbox{CF}_{probe}$ and $\mbox{CF}_{sup}$, respectively.
For simplicity of the analysis, we consider the cases when $ \omega_{probe} \gg \omega_{sup} $ (for LSS) or $ \omega_{probe}\ll \omega_{sup} $ (for HSS), that is, when the probe and suppressor frequencies are not close, as in the case of Figs.~\ref{fig2} and \ref{fig3}.
The configurations of the two oscillators corresponding to LSS and HSS are shown in Figs.~\ref{fig4}A and B, respectively.

First, in the LSS case, both probe and suppressor signals propagate to the probe oscillator with CF$_{probe}$  without significant attenuation.
Therefore, the suppression effect on the probe oscillator with $\mbox{CF}_{probe}$ should be close to that of a single oscillator with $\mbox{CF}_{probe}$ subjected to a superposition of probe and suppressor signals,
\begin{align}
F(t) = fe^{i\omega_{probe}t}+ge^{i\omega_{sup}t}.
\end{align}
The dynamics of	such an oscillator with natural frequency $ \omega_{c} =\omega_{probe}$ is given by
\begin{align}
  \dot{z} = \omega_{probe} [(\mu+i)z-|z|^2z+fe^{i\omega_{probe}t}+ge^{i\omega_{sup}t}],
  \label{single}
\end{align}
where $z(t)$ is the complex amplitude of the oscillator.

We assume that the steady solution to Eq.~(\ref{single}) is given by
\begin{align}
  z(t) = Ae^{i\omega_{probe}t} + Be^{i\omega_{sup}t},
\label{single-response}
\end{align}
where $A$ and $B$ are the response amplitudes to the probe and suppressor signals, respectively.
Plugging this into Eq.~(\ref{single}) and collecting the terms with the same frequencies, we find that the response amplitudes of Eq.~(\ref{single-response}) in the steady state are approximately given as real solutions to the following set of equations:
\begin{align}
-(|\mu|+2B^2)A-A^3+f =0, 
	\label{single-amp-eq1}
\end{align}
\begin{align}
B^6&+2(|\mu|+2A^2)B^4
\cr
&+\{(|\mu|+2A^2)^2 + (1-\Delta \omega_{sup})^2\}B^2 = g^2,
	\label{single-amp-eq2}
\end{align}
where $ \Delta \omega_{sup}=\omega_{sup}/\omega_{probe}$.

Because we have assumed $ \omega_{probe} \gg \omega_{sup} $, the term $ (1-\Delta \omega_{sup} )^2 $ becomes dominant in Eq.~(\ref{single-amp-eq2}), so we approximately obtain $(1-\Delta \omega_{sup})^2B^2\approx g^2$.
Thus, the response amplitude $B$ to the suppressor signal is approximately given by
\begin{align}
  B\approx \frac{g}{|1-\Delta \omega_{sup}|}.
  \label{LSS-suppressor-B}
\end{align}
Plugging Eq.~(\ref{LSS-suppressor-B}) into Eq.~(\ref{single-amp-eq1}), we obtain an approximate equation for the response amplitude $A$ to the probe signal as
\begin{align}
  -\left[ |\mu| +2\left( \frac{g}{1-\Delta \omega_{sup}}\right)^2\right]A-A^3+f \approx 0. 
  \label{LSS-suppressor-A}
\end{align}

Next, we consider the HSS case.
The difference from the LSS case is that the suppressor signal passes through the oscillator with $\mbox{CF}_{sup}$ before reaching the probe oscillator with $\mbox{CF}_{probe}$.
The dynamics of the two oscillators with natural frequencies $ \omega_c =\omega_{sup}$ and $\omega_c = \omega_{probe}$ are given by
\begin{align}
  \dot{z}_{sup} =& \ \omega_{sup} [(\mu+i)z_{sup}-|z_{sup}|^2z_{sup}
  \cr
  &+f_{sup}e^{i\omega_{probe}t}+g_{sup}e^{i\omega_{sup}t}]
  \label{eq_rescale_theo_p},
\end{align}
\begin{align}
  \dot{z}_{probe} =& \ \omega_{probe} [(\mu+i)z_{probe}-|z_{probe}|^2z_{probe}
  \cr
  &+f_{probe}e^{i\omega_{probe}t}+g_{probe}e^{i\omega_{sup}t}]
  \label{eq_rescale_theo_s},
\end{align}
where the inputs to the oscillators are assumed to be
\begin{align}
F_{sup}(t) &= f_{sup}e^{i\omega_{probe}t}+g_{sup}e^{i\omega_{sup}t},
\cr
F_{probe}(t) &= f_{probe}e^{i\omega_{probe}t}+g_{probe}e^{i\omega_{sup}t}.
\end{align}
Here, $f_{sup}$ and $g_{sup}$ are the amplitudes of the probe and suppressor signals received by the oscillator with $\mbox{CF}_{sup}$, and $f_{probe}$ and $g_{probe}$ are the amplitudes of the probe and suppressor signals received by the probe oscillator with $\mbox{CF}_{probe}$.

\begin{figure}[tb]
\centering
    \includegraphics[width=0.85\linewidth,clip]{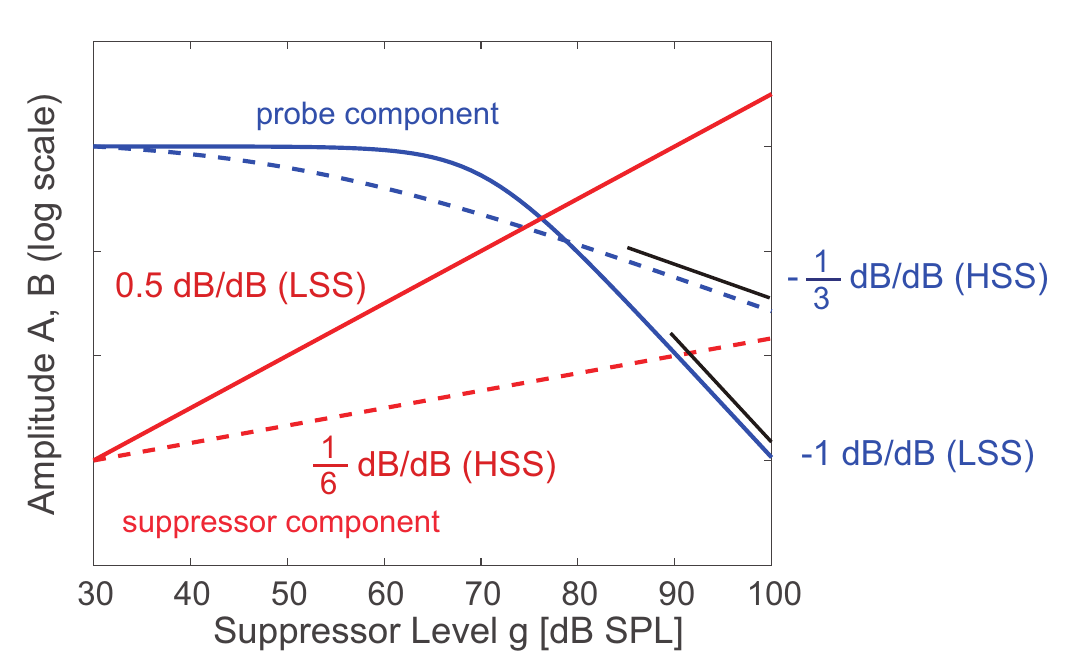}
  \caption{Schematic illustration of the theoretical results for the two-oscillator system, where response amplitudes to the probe and suppressor signals are plotted with respect to the suppressor level in logarithmic scales. Solid and broken lines correspond to the LSS and HSS cases, respectively.}
  \label{fig5}
\end{figure}

As in the previous LSS case, we assume steady response of the oscillators as
\begin{align}
  z_{sup}(t) &= A_{sup}e^{i\omega_{probe}t} + B_{sup}e^{i\omega_{sup}t},
  \label{hss-amplitudes-sup}\\
  z_{probe}(t) &= A_{probe}e^{i\omega_{probe}t} + B_{probe}e^{i\omega_{sup}t},
  \label{hss-amplitudes-probe}
\end{align}
where $A_{sup}$ and $B_{sup}$ are the response amplitudes of the oscillator with $\mbox{CF}_{sup}$ 
to the probe and suppressor signals, and $A_{probe}$ and $B_{probe}$ are the response amplitudes of the oscillator with $\mbox{CF}_{probe}$ to the probe and suppressor signals, respectively.
For the oscillator with $\mbox{CF}_{sup}$, we obtain
\begin{align}
&  A_{sup}^6 + 2(|\mu|+2B_{sup}^2)A_{sup}^4
  \cr
&  +
  \{(|\mu|+2B_{sup}^2)^2+(1-\Delta \omega_{probe})^2\}A_{sup}^2 = f_{sup}^2
\quad
\end{align}
and
\begin{align}
  -(|\mu|+2A_{sup}^2)B_{sup}-B_{sup}^3+g_{sup}=0
\end{align}
by plugging Eq.~(\ref{hss-amplitudes-sup}) into Eq.~(\ref{eq_rescale_theo_p}), where $\Delta \omega_{probe} = \omega_{probe} / \omega_{sup}$ is introduced.
Here, because the response amplitude $ B_{sup} $ to the suppressor signal becomes dominant at the oscillator with $\mbox{CF}_{sup}$, we may approximate the latter equation as
\begin{align}
B_{sup}^3 \approx g_{sup}.
\end{align}
On the other hand, for the oscillator with $\mbox{CF}_{probe}$, we obtain
\begin{align}
  -(|\mu|+2B_{probe}^2)A_{probe}-A_{probe}^3+f_{probe}=0
\end{align}
and
\begin{align}
  &B_{probe}^6+2(|\mu|+2A_{probe}^2)B_{probe}^4
  \cr
  &+\{(|\mu|+2A_{probe}^2)^2+(1-\Delta \omega_{sup})^2\}B_{probe}^2 =g_{probe}^2
\quad
\quad
\quad
\end{align}
by plugging
Eq.~(\ref{hss-amplitudes-probe}) into Eq.~(\ref{eq_rescale_theo_s}),
where $ \Delta \omega_{sup}=\omega_{sup}/\omega_{probe}$ as before.

Because we have assumed $ \omega_{probe} \ll \omega_{sup} $, the term $ (1-\Delta \omega_{sup} )^2 $ is dominant in the above equation and the response amplitude to the suppressor signal at the probe oscillator with $\mbox{CF}_{probe}$ is approximately given by
\begin{align}
  B_{probe}\approx \frac{g_{probe}}{|1-\Delta \omega_{sup}|}, 
\end{align}
and the response amplitude to the probe signal satisfies
\begin{align}
  -\left[ |\mu| +2\left( \frac{g_{probe}}{1-\Delta \omega_{sup}}\right)^2\right]A_{probe}-A_{probe}^3+f_{probe} \approx 0.
\end{align}

\begin{figure}[tb]
\centering
    \includegraphics[width=0.7\linewidth,clip]{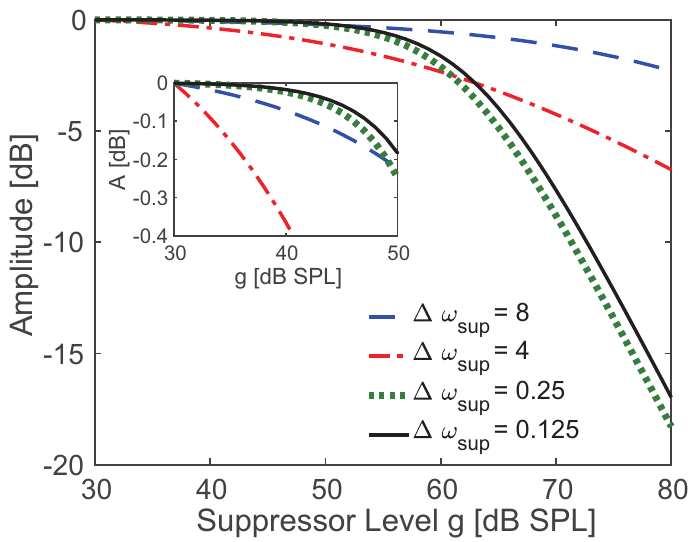}
  \caption{Theoretical curves of the response amplitude to the probe signal vs. suppressor level for the same set of frequency ratios as used in Fig.~\ref{fig3}A. Logarithm of the relative amplitude to the probe signal is plotted as in Fig.~\ref{fig3}A. Results for $\omega_{sup} = 8, 4$ (HSS) and $\omega_{sup} = 0.25, 0.125$ (LSS) are shown.}
  \label{fig6}
\end{figure}

Now, if we assume that the output of the oscillator with $\mbox{CF}_{sup}$ is directly propagated to the oscillator with $\mbox{CF}_{probe}$, that is, if $ f_{probe} = A_{sup}$ and $g_{probe}= B_{sup}$ hold, we obtain 
\begin{align}
  B_{probe} \approx \frac{g_{probe}}{|1-\Delta \omega_{sup}|} \approx \frac{g_{sup}^{\frac{1}{3}}}{|1-\Delta \omega_{sup}|}
  \label{HSS-sup-B}
\end{align}
and
\begin{align}  
  -\left[ |\mu| +2\left( \frac{g_{sup}^{\frac{1}{3}}}{1-\Delta \omega_{sup}}\right)^2\right]A_{probe}-A_{probe}^3+f_{probe} \approx 0,
  \label{HSS-sup-A}
\end{align}
which describe the effect of suppression on the probe oscillator with $\mbox{CF}_{probe}$.

Figure~\ref{fig5} schematically illustrates the above theoretical results for the response amplitudes, where the curves for LSS are given by Eqs.~(\ref{LSS-suppressor-B}) and (\ref{LSS-suppressor-A}), and those for HSS are given by Eqs.~(\ref{HSS-sup-B}) and (\ref{HSS-sup-A}).
It can be seen that the qualitative features of the response curves shown in Fig.~\ref{fig3}A for the probe signal and in Fig.~\ref{fig3}B for the suppressor signal are reproduced by the two-oscillator model.
The curve showing the response amplitude $A_{probe}$ to the probe signal for LSS is steeper than that for HSS and, as far as the approximation in Eqs.~(\ref{LSS-suppressor-B}) and~(\ref{HSS-sup-B}) is valid, the asymptotic slopes of the curves are given by $ -1 $dB/dB (LSS) and $ -\frac{1}{3} $dB/dB (HSS), respectively.
This reflects the different scaling relations of the response amplitude $B_{probe}$ to the suppressor signal on the suppressor amplitude between LSS and HSS, given by Eqs.~(\ref{LSS-suppressor-B}) and~(\ref{HSS-sup-B}).
The asymptotic slope $-1$ for LSS is close to the slope $ -0.9 $dB/dB obtained experimentally by Rhode {\it et al.}~\cite{Rhode1993}, and the result for HSS also qualitatively agree with the experimental result in that the slope is shallower than the LSS case.

Figure~\ref{fig6} shows theoretical curves of the response amplitude $A_{probe}$ to the probe signal for several frequency ratios used in the numerical simulations shown in Fig.~\ref{fig3}A.
As in Fig.~\ref{fig2}, logarithm of relative response amplitudes to the probe signal, $10 \log_{10} (A_{probe} / A_{probe, 30{\rm dB}})$, is plotted.
The theoretical curves for the two-oscillator system reproduce the results of numerical simulations for the $10$-oscillator system in Fig.~\ref{fig3}A qualitatively well. That is, the suppression is stronger when the suppressor frequency is closer to the probe frequency, and the decay of the curves for HSS is much slower than those for LSS.

\begin{figure}[h]
\centering
    \includegraphics[width=1\linewidth,clip]{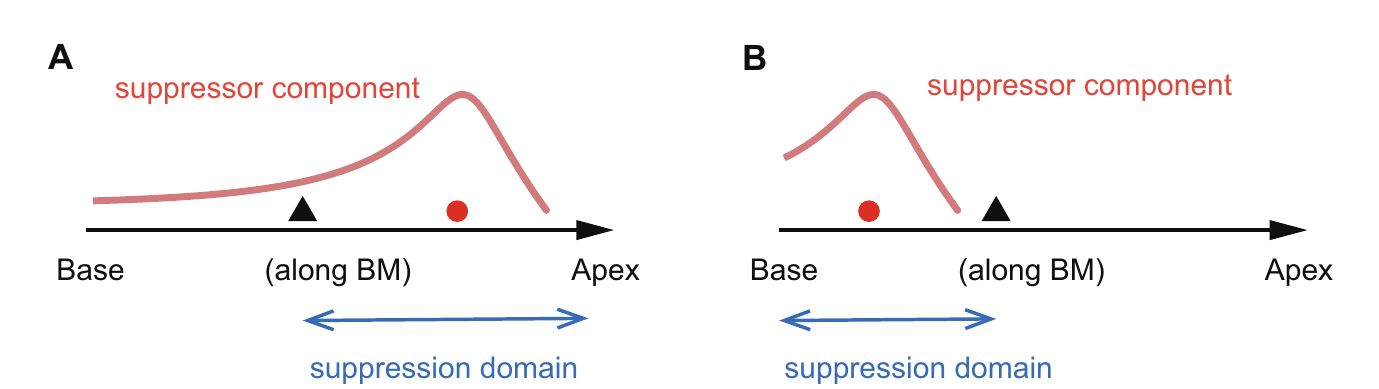}
  \caption{Schematic illustration of the configuration of the oscillators with $\mbox{CF}_{probe}$ and CF$ _{sup}$ and the suppression domain. (A) LSS, (B) HSS.
 }
  \label{fig7}
\end{figure}

The above theoretical results suggest that the nonlinear amplification and compression property in the cochlea can give rise to qualitatively different suppression properties between HSS and LSS.
Figure~\ref{fig7} schematically illustrates the difference in the configuration of the two oscillators with $\mbox{CF}_{probe}$ and $\mbox{CF}_{sup}$ between LSS and HSS, as well as the propagation of the suppressor signal and the suppressed domain.
In the LSS case, the suppressor signal propagates through the probe oscillator with $\mbox{CF}_{probe}$ without significant amplification and then later amplified near the oscillator with $\mbox{CF}_{sup}$. Thus, the suppression domain arises behind the oscillator with $\mbox{CF}_{probe}$ as shown in Fig.~\ref{fig2}A.
In contrast, in the HSS case, the suppressor signal is nonlinearly amplified before reaching the probe oscillator with $\mbox{CF}_{probe}$. Thus, the suppression domain arises in front of the oscillator with $\mbox{CF}_{probe}$ as shown in Fig.~\ref{fig2}B.
Moreover, because of the nonlinear amplification and compression of the suppressor signal in the HSS case, the response amplitude exhibits much slower decay with the increase in the suppressor level than that in the LSS case
as shown in Fig.~\ref{fig3}A.

\vspace*{-3mm}
\section{Summary}

Using a one-dimensional array of Stuart-Landau oscillators with feed-forward coupling as a model of the cochlea, we have studied two-tone suppression effect by analyzing dependence of the response amplitude to the probe signal on the frequency and level of the suppressor signal.
We have found by numerical simulations that the suppression effect is qualitatively different between the HSS and LSS cases, which is also observed in physiological experiments.
By theoretically analyzing a simplified two-oscillator model, we have clarified that the difference between HSS and LSS is caused by the difference in the relative configuration of the oscillators with $\mbox{CF}_{probe}$ and $\mbox{CF}_{sup}$. In particular, the nonlinear amplification and compression property plays an important role in the case of HSS.

The discrepancies of the HSS curves between the numerical simulations and theoretical analysis are caused by the simplification of the original $10$-oscillator system to a two-oscillator system.
This can be improved by considering more complex models, but the essential reason for the qualitative difference between LSS and HSS is already clear from the present theoretical analysis on the two-oscillator system.
The insights gained in this study may be relevant in understanding the auditory mechanism  of two-tone suppression phenomena in the actual cochlea.

\vspace*{-3mm}
\acknowledgments

We thank Professor Tohru Kohda for drawing our attention to coupled-oscillator models of the cochlea. We also thank financial support from JSPS (Japan), KAKENHI Grant Numbers JP16H01538, JP16K13847, and JP17H03279.

\end{document}